\pdfoutput=1
\documentclass[sigconf]{acmart}

\AtBeginDocument{%
  }

\copyrightyear{2025}
\acmYear{2025}
\setcopyright{acmlicensed}\acmConference[SIGCSE TS 2025]{Proceedings of the 56th ACM Technical Symposium on Computer Science Education V. 1}{February 26-March 1, 2025}{Pittsburgh, PA, USA}
\acmBooktitle{Proceedings of the 56th ACM Technical Symposium on Computer Science Education V. 1 (SIGCSE TS 2025), February 26-March 1, 2025, Pittsburgh, PA, USA}
\acmDOI{10.1145/3641554.3701870}
\acmISBN{979-8-4007-0531-1/25/02}

\usepackage{balance}

\begin{document}

\title[Teaching Cloud Infrastructure in an Undergraduate Computer Science Program]{Teaching Cloud Infrastructure and Scalable Application Deployment in an Undergraduate Computer Science Program}


\author{Aditya Saligrama}
\email{saligrama@cs.stanford.edu}
\affiliation{%
  \institution{Stanford University}
  \city{Stanford}
  \state{California}
  \country{USA}}

\author{Cody Ho}
\email{codyho@cs.stanford.edu}
\affiliation{%
  \institution{Stanford University}
  \city{Stanford}
  \state{California}
  \country{USA}}

\author{Benjamin Tripp}
\email{btripp@cs.stanford.edu}
\affiliation{%
  \institution{Stanford University}
  \city{Stanford}
  \state{California}
  \country{USA}}

\author{Michael Abbott}
\email{michael.abbott@acm.org}
\affiliation{%
  \institution{Stanford University}
  \city{Stanford}
  \state{California}
  \country{USA}}

\author{Christos Kozyrakis}
\email{christos@cs.stanford.edu}
\affiliation{%
  \institution{Stanford University}
  \city{Stanford}
  \state{California}
  \country{USA}}

\renewcommand{\shortauthors}{Aditya Saligrama, Cody Ho, Benjamin Tripp, Michael Abbott, \& Christos Kozyrakis}

\begin{abstract}
  Making successful use of cloud computing requires nuanced approaches to both system design and deployment methodology, involving reasoning about the elasticity, cost, and security models of cloud services.
Building cloud-native applications without a firm understanding of the fundamentals of cloud engineering can leave students susceptible to cost and security pitfalls.
Yet, cloud computing is not commonly taught at the undergraduate level.
To address this gap, we designed an undergraduate-level course that frames cloud infrastructure deployment as a software engineering practice.
Our course featured a number of hands-on assignments that gave students experience with modern, best-practice concepts and tools including infrastructure-as-code (IaC).
We describe the design of the course, our experience teaching its initial offering, and provide our reflections on what worked well and potential areas for improvement.
Our course material is available at \underline{\url{https://infracourse.cloud}}.

\end{abstract}

\ccsdesc[500]{Social and professional topics~Computing education}
\ccsdesc[500]{Software and its engineering~Cloud computing}

\keywords{cloud computing; application deployment; scalability; infrastructure-as-code; computing education}

\maketitle

\section{Introduction}

Modern organizations are increasingly relying on cloud infrastructure for their computing workloads to better scale applications or run high-performance experiments.

Making optimal use of cloud infrastructure is a nontrivial engineering problem.
Cloud-native development and deployment diverges from traditional, on-premise practices in a number of key ways that together represent a paradigm shift in the mindset required to approach such tasks.
For example, modern cloud-deployed applications often run on compute platforms where state is ephemeral and thus must be designed to store state in separate managed services \cite{McCance2012CERN}.
Cloud infrastructure gives developers tools to more finely control compute and storage capacity needed to support their applications \cite{Lehrig2015Scalability}.
Securing cloud-deployed applications also requires understanding a different set of abstractions.
Though such workloads run on third-party hardware, cloud providers only assume security responsibility for the cloud platform, while delegating the responsibility of how cloud services are used to developers \cite{AWS2024SharedResponsibility,Google2023SharedResponsibility,Azure2023SharedResponsibility}.

Effectively and securely working with cloud platforms is tricky and requires a nuanced understanding of cloud-native application design and deployment concepts.
Many engineers experience difficulties stemming from a poor understanding of security, elasticity, and maintenance, with potentially severe costs.
For example, in late 2023, attackers exploited leaked AWS API keys to mine cryptocurrency within victims' AWS accounts, leading to thousands of dollars in unexpected costs \cite{McCloskey2024CryptoMining}.
Other issues can stem from choosing the wrong data storage service, which can lead to poor scalability and high data transfer expenses.

Students often encounter cloud abstractions early in their careers, both within the university context and in industry.
Many CS curricula in universities, including our institution, feature capstone projects where students work cooperatively on a complex web application \cite{Li2023Capstone,Shakil2024Capstone}, where deployment is often a required component.
Instruction on doing so is typically limited and relegated to the end of the course, mostly focusing on deploying to a single virtual machine (VM) rather than meaningfully interacting with modern, managed cloud abstractions.
Similarly, during software engineering internships, students often work with cloud-deployed services such as containers in clusters or deploy new services themselves.

Effectively preparing students for these academic and industry roles requires a more comprehensive primer on working with cloud services.
Although there are a number of existing courses that cover cloud computing, most are offered at a graduate or professional level and focus largely on the engineering and research behind cloud platforms themselves \cite{BerkeleyCloudCourse,StanfordCloudCourse,CornellCloudCourse}, rather than on deploying and designing cloud-native applications.
Cloud deployment is not commonly taught at the undergraduate level, as cloud infrastructure offerings went through a period of rapid change in the past and often diverged drastically between providers \cite{Kroonenburg2020Cloud}.
But, in recent years, there has been a maturation of the technologies and practices involved in deployment beyond provider-specific tools.
These concepts, such as infrastructure-as-code (IaC), containerization and container orchestration, functions-as-a-service (FaaS), and other managed abstractions on top of open-source compute and data services, are also now more relevant within a software engineering career as the separation between engineering and operations has narrowed \cite{Buchanan2024DevOps}.

Accordingly, we developed a new undergraduate course known as "Cloud Infrastructure and Scalable Application Deployment," which focuses on building student fluency with these cloud infrastructure as a service (IaaS) concepts and tools in a hands-on manner.
Our course consists of lectures that emphasize application design in a way that best takes advantage of modern cloud services.
Our course additionally offered students a safe sandbox in which to experiment with cloud deployments, while providing structure in learning how to do so systematically.
This involved designing a novel set of assignments that gave students practice in deploying applications in modern patterns using IaC.

We present insights from our experience teaching the inaugural offering of the course in Winter 2024.
We hope that our reflections and course materials can be useful to educators at other institutions.

\section{Related Work}
The need for universities to teach cloud computing skills has been recognized for over a decade \cite{Border2013Cloud}. However, most courses are offered by continuing education programs, targeting professionals who are looking to reskill into cloud computing and DevOps careers \cite{UTPostGradCloudCourse,CaltechCloudCourse}.
Often, the focus of these courses is to prepare their students for industry certifications offered by cloud providers.
These courses generally do not expect any CS background from their students, and as such, they do not offer much in the way of technical assignments.

There are also a number of cloud computing courses that are offered as part of undergraduate or graduate CS programs. 
Broadly, we find two categories of such courses.
Courses in the first category focus on application-level concepts and tools and use the cloud as a deployment vehicle (typically to easily spin up VMs), but do not fully engage with more cloud-native abstractions and managed services \cite{DecalCloudComputing,Liang2018WebDev}.
Courses in the second category focus on the engineering of cloud platforms themselves, either through teaching the foundational technologies that power cloud systems such as distributed systems \cite{CornellCloudCourse} or through surveying recent research papers in the field \cite{StanfordCloudCourse,BerkeleyCloudCourse}.
But these courses are not accessible to undergraduate students without substantial prior systems background, and do not address the need for such students to learn to build systems that effectively interface with cloud services.

Usage of the cloud also features in a number of other CS courses, even if it is not the main focus.
Many universities offer courses on DevOps, in which students learn about the tools and practices that enable rapid software development and deployment \cite{Benni2019DevOps,Alves2021DevOps,Fernandes2022DevOps,Verdicchio2023DevOps}.
Cloud infrastructure is a key part of DevOps, as it enables the automation of pipelines for building, testing, and deploying software.
However, these courses mostly focus on the processes that make such a software lifecycle efficient, such as continuous integration and test orchestration, rather than the infrastructure that supports both these processes and the deployed applications.

Some recent DevOps courses do engage with more cloud-native concepts and tools.
\citet*{Christensen2016DevOps,Christensen2022DevOps} designed a DevOps course that emphasized application and infrastructure design for scalability and reliability, similar to our course, and focused on refactoring and containerizing an application into a microservices pattern.
\citet*{Demchenko2019Cloud} taught a course on DevOps and cloud-based software development that, like us, incorporated the use of IaC tools such as Terraform, AWS CloudFormation, and Ansible.
But these courses do not give equal weight to cloud-native application design and the cloud services used in their deployment.
These skills are equally important in building systems that interact with cloud abstractions, and effectively doing so requires a strong integrated understanding of both concepts \cite{Garrison2017CloudNative}.

Similarly, \citet*{Dickerson2023SRE} recently designed and taught a course focusing on site reliability engineering (SRE) principles.
The SRE discipline focuses on maintaining the reliability of deployed production systems, many of which run on cloud infrastructure.
Their course made use of some of the same tools as our course, including AWS services for compute, networking, and storage, but focused primarily on the operational aspects of maintaining a service in production rather than its initial design and deployment.

\begin{table*}[t]
  \caption{Content covered in lectures. Guest lectures are indicated by asterisk (*).}
  \label{tab:lectures}
  \begin{tabular}{lll}
    \toprule
    \textbf{Week} & \textbf{Lecture 1} & \textbf{Lecture 2} \\
    \midrule
    1 & Building Blocks of Cloud Infrastructure & Networking Primer; Web App Primer \\
    2 & \textit{Holiday, no lecture} & Cloud Networking (VPC, Load Balancing, CDN) \\
    3 & Cloud Storage (Object Storage, Databases, Caches) & Database Tradeoffs for Cloud Applications* \\
    4 & Containerization and Container Orchestration & Infrastructure-as-Code and Cloud Automation \\
    5 & Identity and Access Management; Cloud Security Topics & Auditing, Logging, and Observability \\
    6 & Serverless Computing and FaaS & Cloud Machine Learning and High-Performance Computing \\
    7 & \textit{Holiday, no lecture} & Continuous Integration and Continuous Deployment \\
    8 & Scaling Applications in Industry* & Practical Considerations in Cloud-Native Application Design \\
    9 & Ethical Issues in Cloud Computing & Cloud Billing and Cost Management* \\
    10 & Evolution of the Cloud Ecosystem and Cloud-Native Design* & Recap and Misc Topics \\
    \bottomrule
  \end{tabular}
\end{table*}

\section{Course Overview and Logistics}

We designed a 10-week course entitled "Cloud Infrastructure and Scalable Application Deployment" and offered it as an undergraduate elective at Stanford University in Winter 2024.

\subsection{Learning Objectives}

By the end of the course, students should be able to:

\begin{enumerate}
    \item Design cloud-native applications to take advantage of the elasticity, cost, and system administration benefits provided by cloud infrastructure.
    \item Architect a cloud deployment by selecting and combining appropriate compute, networking, and storage resources.
    \item Systematically and reliably deploy cloud resources using IaC.
    \item Ensure that cloud deployments remain secure, observable, and continuously updated.
\end{enumerate}

\subsection{Class Composition}

In total, 50 students were enrolled in the course, including 15 women and 7 students from underrepresented minority backgrounds.
Most students were either juniors, seniors, or graduate students, although there were a limited number of underclassmen enrolled.
Although the course was open to all students, most students were CS majors.

We intentionally required only programming maturity up to an introductory systems class and some experience using command-line tools in order to keep the course accessible to underclassmen.
However, of the 34 students that completed an intro survey and remained enrolled in the class until the end, 76\% had completed an operating systems course and 41\% had completed a web or mobile development course.
Additionally, 79\% had some experience with backend or frontend web development, 68\% had experience using an IaaS cloud provider, and 35\% had experience with containerization.
The first offering of the course was a pilot that more experienced students likely self-selected into, and we expect that future offerings will have a more diverse set of students.

\subsection{Course Structure}

Over 10 weeks, the course met for 80 minutes twice a week.
Lectures covered the core areas of cloud application deployment, detailing the types of services offered by public cloud providers and the considerations behind selecting and deploying these services for a given application.
We additionally discussed system and application design for scalability, security, and cost efficiency.

There were four hands-on programming assignments, each worth 15\% of a student's final grade, which we describe in Section~\ref{sec:assignments}.
Assignments focused on systematically deploying a pre-built application to the cloud in modern best-practice patterns using IaC.
We encouraged students to work in pairs on these assignments, and students were allowed to discuss conceptual strategies with other students in the class.

Lastly, the course featured a final project, worth 40\% of students' final grades, where students were given the opportunity to deploy an application of their own choosing to the cloud.

\section{Course Content}

Our course presented fundamental practices necessary for publishing and managing usable web applications on the public cloud as well as application design that takes advantage of cloud abstractions.
Our lectures, assignments, and infrastructure for managing the class are open-source and publicly available online at \underline{\href{https://infracourse.cloud}{https://infracourse.cloud}}.

\subsection{Lecture Content}

The first half of our course discussed the building blocks of cloud infrastructure that help in deploying web applications, including the basics of compute, network, and storage resources on the cloud, containers and container orchestration, and serverless and edge computing using Functions-as-a-Service (FaaS).

We then shifted our focus to patterns in which these services can be used to deploy scalable applications.
This included coverage of auditing, logging, and observability; the basics of continuous integration and continuous deployment; and machine learning and high-performance computing on the cloud.
Additionally, we discussed using software development kits (SDKs) to let applications interact with their underlying infrastructure in order to (e.g.) fetch objects from storage buckets.

Security was a significant focus area.
We emphasized the shared responsibility model of cloud security, where providers take responsibility for securing the hardware and software underlying cloud platforms whereas developers take responsibility for how they use cloud services.
We included a dedicated lecture on identity and access management (IAM) and its high attack surface in the context of modern serverless and containerized workloads running on the cloud.
For all other lectures, where relevant, we included security considerations and best practices to minimize the attack surface of the services we discussed.

We covered all conceptual material and general concepts in lectures in a cloud provider-agnostic fashion, but demonstrated specific examples on Amazon Web Services (AWS) where required.
We summarize the content covered in these lectures in Table~\ref{tab:lectures}.

\subsubsection{Guest Lectures}

We invited four guest lecturers from industry to provide insight and depth on select topics.
All speakers were involved with leading the design and deployment of cloud-native applications at their respective employers.
We summarize the content they covered in Table~\ref{tab:lectures}.

\subsection{Technical Assignments}\label{sec:assignments}

\begin{table}[h]
    \caption{Technologies exercised in each assignment}
    \label{tab:assignment_technologies}
    \begin{tabular}{ll}
        \toprule
        \# & \textbf{Technologies exercised} \\
        \midrule
        1 & VMs, Web Application Deployments, DNS and TLS \\
        2 & IaC, IAM, Networking, Databases, Object Storage, Containers \\
        3 & Observability, Functions-as-a-Service (FaaS) \\
        4 & Continuous Integration \& Deployment, Container Registries \\
        \bottomrule
    \end{tabular}
\end{table}

Our course featured four hands-on technical assignments that were designed to give students experience with the cloud tools and services discussed in lectures.
Our assignments focused on deploying web services to AWS, which we chose due to its predominant market share and the maturity of its cloud service offerings.
AWS, via our institution's account manager, additionally granted each student \$200 in credits for the duration of the course to cover resource costs.
However, we emphasize that these assignments can be generalized to any cloud provider.

The first assignment was a simple exercise in deploying a static website to the cloud.
Here, students created an EC2 VM running Ubuntu Linux, installed the Nginx web server software, added a static web page, and configured Let's Encrypt to add support for HTTPS.
This was intended to refresh students' familiarity with the command line, as well as to set up their AWS accounts and to introduce them to working with the AWS console.
Additionally, we wanted students to get a sense of what technologies are used under the hood of cloud deployments, before adding in the abstractions of managed services.

Assignments 2 through 4 focused on building up a deployment of a more full-featured web application with a frontend and backend.
For this purpose, we designed a custom, minimal image-sharing social media platform called \textit{Yoctogram} written in Python (FastAPI) and React backed by the PostgreSQL database.
We chose this stack due to its popularity and similarity to real web applications students are likely to encounter.

For Assignment 2, students created the initial deployment of Yoctogram to AWS using IaC.
We chose to use the AWS Cloud Development Kit (CDK) for this assignment, as it allowed students to define their infrastructure in Python, with which they were already familiar.
Students wrote CDK constructs to bring up network (VPC, public and private subnets, DNS via Route 53, TLS via Certificate Manager, Cloudfront CDN), storage (Aurora Postgres Serverless database, S3 object storage), and compute (ECS Fargate container cluster) services to support Yoctogram.
This was the course's most difficult and time-consuming assignment, given the broad scope of the resources students needed to define and the difficulty in debugging the deployment, which we discuss further in Section~\ref{sec:debugging}.

Assignment 3 focused on adding observability to the Yoctogram deployment, as well as interfacing with FaaS via AWS Lambda to add an image compression feature to Yoctogram.
Students configured a Datadog free trial account and added a sidecar container definition in CDK in order to send Yoctogram's backend container logs to Datadog.
Students also added CDK code to instantiate a preprovided Lambda function for image compression that operated on image-upload S3 buckets.
We verified that students had correctly set up Datadog by sending their Yoctogram deployment an intentionally failing web request, which they needed to find in Datadog and report a unique "flag" we set in the \texttt{User-Agent} HTTP header.

Lastly, Assignment 4 involved creating a continuous integration and deployment (CI/CD) pipeline for Yoctogram in order to automate the deployment process on code changes.
Students wrote a GitHub Action that triggered on changes to the \texttt{main} branch of their Yoctogram repository in order to build a Docker container and tag it in Elastic Container Registry for release to production.

All student deployments were expected to follow the cloud security best practices discussed in lectures.
For example, students were expected to use IAM roles and policies to restrict access to their resources, encrypt data at rest and in transit, and use network segmentation to isolate their resources.
We summarize the technologies exercised in each assignment in Table~\ref{tab:assignment_technologies}.

Student completion rates were high for each assignment: 100\% completed Assignment 1, 92\% completed Assignments 2 and 3, and 90\% completed Assignment 4.

\subsection{Final Project}

The final project tasked students with deploying a web application of their choosing to the public cloud.
Students were allowed to choose either an application they designed themselves or an existing open-source project.
The project was a way for students to cumulatively apply the concepts they learned without the scaffolding we provided in the assignments.
In particular, it gave students the opportunity to \textit{architect} a new deployment rather than merely instantiate cloud resources for an existing architecture.

Students were expected to tie together multiple cloud services using IaC to create an elastic deployment that could scale with demand and followed security best practices.
Deliverables included the source code for the IaC used to deploy the application, a brief write-up of the deployment's architecture and services used, and a link to the deployed application for evaluation.

Many students demonstrated substantial creativity with the final project, illustrating their command of course concepts.
One group deployed an arcade Minecraft server that ran on the ECS Fargate container orchestration platform with a CI/CD pipeline to build game plugins, such as a stateful economy modification that stored data in a managed Aurora MySQL database.
Another student built a Chrome extension connected to a backend they deployed on ECS Fargate that summarized lecture videos into concise notes using large language models.

\section{Discussion}

Overall, students enjoyed the course and found it to be a valuable learning experience.
We draw on feedback from an end-of-quarter evaluation survey conducted by our institution, though we did not conduct a rigorous evaluation of the course.
The survey asked students to score how well they thought they had achieved our learning goals, how much they learned from the course, how well the course was organized, and how they would rate our quality of instruction.
Additionally, students were also given the opportunity to describe the skills they learned in the course, and recommendations they had for students taking the course in the future.

In total, 39 students filled out the survey.
72\% of students stated they learned ``a great deal'' or ``a lot'' (a score of 4 or 5 on a 1-5 scale).
Students broadly found value in the practical skills they learned: a few students reported that the course built up ``actual confidence in my ability to use the cloud intelligently,'' and that ``this was the missing building block in building full-stack cloud awareness.''
Students also overwhelmingly reported that they found the course useful for their future careers.

We also received feedback that the course was harder and more time-consuming than students expected.
We believe this was partially due to the number of new concepts, services, and tools that students were exposed to, as well as general challenges common to any first offering of a course.
However, many issues also stemmed from our miscalibration of assignment difficulty with the time we gave students to complete them, the scaffolding we provided in assignment starter code that prevented students from making more architectural choices, and state management issues with the AWS CDK that led to significant challenges in debugging deployments.
We are currently in the process of addressing these issues based on student feedback for the next offering of the course.

In the following subsections, we provide insights on the first offering of our course based on our own reflections.

\subsection{Successes}

\subsubsection{Course infrastructure}
To easily manage a course of 50 students with only three staff, we built a significant amount of custom infrastructure to support students given the variety of resources they interacted with throughout the course.

In particular, we observed a need to allow students to provision themselves certain necessary resources in a self-service manner without manual intervention from course staff.
Towards this end, we built lightweight tooling to provision resources like DNS records and AWS credits.
We also provided students with an AWS-hosted VM template for their development environment to eliminate potential dependency issues on their own machines.

Grading student submissions, for which we built custom autograders for each assignment, was also a complex task as we needed to check both their IaC and the frontend and backend components of their deployed application.
To verify student IaC conformance to our specifications and cloud security best practices, we wrote a set of Open Policy Agent (OPA) checks as part of the autograder.
We also orchestrated a Chromium browser to run a headless test suite against a student's deployed web application.
We found that our autograders were robust in handling student submission code; in no case did these autograders assign students undeserved points.

Additionally, we also distributed the OPA checks to students alongside assignments such that they could be run locally and provide immediate feedback.
This helped us add guardrails to let students build and experiment while keeping them safe from common security pitfalls and deployment errors without substantial pedagogical value.
Future improvements include more detailed submission feedback on student IaC mistakes and deployment bugs in order to help students troubleshoot more effectively.


Our infrastructure was built using the same cloud-native best practices central to the course.
This allowed us to feature it in lecture demos, which further illustrated to students practical examples of the concepts they were learning.
Our code is open-source.

We plan to further extend our infrastructure for the next offering.
We might add an IaC provider for our custom provisioner to reduce manual intervention during student deployments.
We may also add performance testing to our autograders to ensure that students' deployments can handle a certain amount of load.

\subsubsection{Course scope for target audience}
The course's framing was well suited towards its target audience of undergraduate students interested in future technical software careers.
Our course integrated cloud computing concepts and application design principles towards architecting and managing cloud-native systems, thus treating infrastructure deployment as a software engineering practice.
Many students reported that this course structure helped address the academia-industry gap through applying fundamental CS concepts they learned in other courses to ours.

Additionally, our guest speakers provided valuable lessons that drove course material forwards by engaging in-depth with how they interface with and design applications on top of cloud infrastructure in industry.
The class's overall participation and engagement with guest speakers was high.
Students particuarly enjoyed the lecture on cloud billing as it helped vindicate their frustration with some of the sharp edges they encountered with AWS during their technical assignments.
Many guest speakers independently commented on how insightful students' questions were, illustrating how they enhanced students' understanding of course material.

\subsubsection{IaC everywhere}
Assignments 2-4 involved having students write IaC in AWS CDK to deploy infrastructure.
One benefit of this approach was that it helped students better conceptualize infrastructure deployment as a software engineering process by allowing them to better visualize cloud dependencies between resources through writing code in a language they were familiar with.

IaC also helped students more easily troubleshoot their deployments.
When deployments failed, they could more easily compare their code against the assignment requirements and identify mistakes than if they had to manually click through the AWS console to check their resources.
Moreover, if student mistakes required a complete redeployment of their infrastructure, they could do so with a single command, saving them time and frustration.

While centering assignments around IaC was a success, the specific choice of IaC flavor in CDK created state management challenges.
We found that when students made mistakes in their code, CDK would not always provide clear error messages or roll back the deployment in a timely manner (sometimes taking over an hour), which led to student frustration and required us to provide significant hands-on debugging help during office hours.
We are evaluating other IaC options for the next offering of the course, such as OpenTofu, which does not suffer from the same issues.

\subsection{Challenges and Areas for Improvement}

\subsubsection{Building domain-specific debugging skills}\label{sec:debugging}
Effectively debugging software draws on testing the system's behavior against falsifiable hypotheses formed from the system's design and the observed behavior \cite{Fitzgerald2010Debugging, ODell2017Debugging}.
Within the context of cloud infrastructure, the process of forming such hypotheses requires a strong operational understanding of both the cloud services being used and the application being deployed.
For example, if a web application does not load properly, a student must check server logs for application errors, then check the network configuration to ensure the server is reachable from the internet, and then check the CDN configuration to ensure it has permissions to cache the server's content.

But there are a number of sharp edges in cloud services that make gathering this information difficult.
Many cloud services do not persist logs by default, and even when they do, it can be difficult to access, search, and visualize them.
Error messages can be cryptic and unhelpful, and sometimes services will fail silently without providing any feedback.
Moreover, this debugging model presumes a low iteration time while testing hypotheses, but cloud services can take minutes to hours to deploy, making debugging a slow and frustrating process.
This was a significant challenge for students, and often required debugging intervention from course staff.

Although we obviously cannot eliminate these inherent issues with cloud services, we can better prepare students to debug their deployments.
We plan to provide students with more written material on finding useful debugging signals such as logs and accessibility analytics, and provide preconfigured constructs in assignment starter code to ensure that these signals are available to students.
We also plan to expand our static test suite to help students catch more classes of deployment errors that are difficult to debug yet not pedagogically relevant, such as typos in environment variable names.

\subsubsection{Architectural freedom vs. course scaling}
Our lecture material strongly emphasized application-level considerations for cloud-native design.
However, we are continuing to iterate on how best to let students make architectural decisions in technical assignments.

One consideration is that architectural freedom introduces heterogeneity in the types of solutions that students work towards.
This sometimes causes different students to head down "paths that were impossibly long without realizing that the time and effort they were spending was not reasonable given the task" \cite{Dickerson2023SRE}.
It also makes autograding more difficult.
This creates additional work for course staff in keeping students away from avoidable rabbit holes and in grading submissions, which limits the scale of the course.

During our pilot offering, we erred on the side of providing students with more scaffolding for IaC deployments.
Unfortunately, this made it more difficult for students to practically apply application design concepts they learned in lecture.
Students did interface with such concepts in instrumenting applications for observability and in some final projects that featured custom application logic.
However, we recognize that we could have done more to let students explore these concepts in their assignments.

In the future, we intend to give students more architectural freedom for their submissions.
For example, we can have students build further features for Yoctogram that take advantage of cloud-native constructs, such as image manipulation using functions-as-a-service.
We can also envision grading some assignments based on performance and reliability rather than on the IaC itself, which may help in making assignments more open-ended without much additional overhead.

\subsubsection{Adapting to students' backgrounds}
Cloud deployment draws on a number of fundamental CS concepts, such as in networking and web systems.
However, we did not require students to have any prior experience with these topics.
Students at our institution typically take such courses as upperclassmen, but we wanted the course to be accessible to younger students, many of whom already interface with cloud services outside coursework.
Moreover, at our institution, these courses often explore their subject area more deeply than necessary for our material; for example, our networking course has students implement the TCP/IP stack.

To address this gap, we included primer lectures on computer networking and web applications early in the term.
Although these lectures were well-received, we still found that less experienced students encountered some difficulty with troubleshooting inaccessible deployments.
This stemmed from a lack of intuition surrounding the network and web processes in play when loading a web application.
We found ourselves needing to repeatedly recap some of these concepts in later lectures and during office hours.

In the future, we plan to provide more written material on these topics and strengthen the primer lectures to better bring students up to speed with these topics.
That being said, educators at other institutions may make other decisions with regards to setting prerequisites.
Where courses that deal with web systems and networking are typically taken earlier in student careers, it may make more sense to simply set these courses as a prerequisite and preserve more lecture time for cloud computing content.

\subsubsection{Account and billing logistics}
Students worked on technical assignments using accounts they created themselves into which they could deposit the \$200 credit codes we provided.
This required students to enter credit card information as a backup in case of overspend, but allowed students to keep unused credits after the term to use on personal projects.

The vast majority of students were able to avoid overspend as completing the assignments and final project in pairs allowed them to use a second AWS account in case funds in one account ran low.
However, a few students did not properly clean up resources after the end of the term and thus saw charges on their credit cards.
In these cases, we were able to work with AWS to waive the charges and get the resources shut down.

This process could be improved by having our institution provision student accounts using AWS Organizations and IAM Identity Center, which transfers overspend liability from the student to the institution.
Managing such infrastructure would increase the course staff's workload, but it prevents students from incurring charges and would also allow staff to more easily troubleshoot student account issues.
We also plan to cover basic billing topics earlier in the course and to ensure students set billing limits on their accounts to prevent overspend.

\section{Conclusion}

In this work, we presented our experience teaching an undergraduate course on cloud infrastructure and scalable application deployment.
Students learned practical skills related to deploying and managing cloud infrastructure, as well as application design that takes advantage of managed cloud services.
Our course featured a number of hands-on assignments that gave students experience with modern, best-practice concepts such as IaC, containerization, and serverless computing.
While there are several improvements we can make to future offerings, both students and course staff felt that the course was successful in building student fluency with cloud fundamentals and tools useful for launching scalable applications and for further career growth.
We hope that our experiences and public course materials can be useful to educators interested in addressing the gap in practical cloud computing education for undergraduate students.

\section*{Acknowledgements}


We thank Zakir Durumeric and Kimberly Ruth for advice on the design of the course and Veronica Rivera for feedback on this paper.

\bibliographystyle{ACM-Reference-Format}
\balance
\bibliography{main}

\end{document}